\documentclass[showpacs,amsmath,amssymb,amsfonts,pre,floatfix]{revtex4} 
\usepackage{graphicx}
\usepackage{bm}
\begin{document}

\title{Scattering off an oscillating target: Basic mechanisms and their impact on cross sections}

\author{I. Brouzos}\email{brouzos@physi.uni-heidelberg.de}
\affiliation{Department of Physics, University of Athens, GR-15771
Athens, Greece}

\author{A.K. Karlis}\email{alkarlis@med.uoa.gr}
\affiliation{Department of Physics, University of Athens, GR-15771
Athens, Greece}

\author{C. A. Chrysanthakopoulos}
\affiliation{Department of Physics, University of Athens, GR-15771
Athens, Greece}

\author{V. Constantoudis}
\email{vconst@imel.demokritos.gr}\affiliation{Institute of
Microelectronics, NCSR Demokritos, P.O. Box 60228, Attiki, Greece}

\author{F.K. Diakonos}
\email{fdiakono@phys.uoa.gr} \affiliation{Department of Physics,
University of Athens, GR-15771 Athens, Greece}

\author{P. Schmelcher}
\email{Peter.Schmelcher@pci.uni-heidelberg.de}
\affiliation{Physikalisches Institut, Universit\"at Heidelberg,
Philosophenweg 12, 69120 Heidelberg, Germany}
\affiliation{Theoretische Chemie, Im Neuenheimer Feld 229,
Universit\"at Heidelberg, 69120 Heidelberg, Germany}

\author{L. Benet}
\email{benet@fis.unam.mx} \affiliation{Instituto de Ciencias Fisicas,
Universidad Nacional Autonoma de Mexico (UNAM), Apdo. Postal 48-3, 62251 Cuernavaca, Mexico}

\date{\today}

\begin{abstract}
We investigate classical scattering off a harmonically oscillating target in two spatial dimensions. The shape of the scatterer is assumed to have a boundary which is locally convex at any point and does not support the presence of any periodic orbits in the corresponding dynamics. As a simple example we consider the scattering of a beam of non-interacting particles off a circular hard scatterer. The performed analysis is focused on experimentally accessible quantities, characterizing the system, like the differential cross sections in the outgoing angle and velocity.
Despite the absence of periodic orbits and their manifolds in the dynamics, we show that the cross sections acquire rich and multiple structure when the velocity of the particles in the beam becomes of the same order of magnitude as the maximum velocity of the oscillating target. The underlying dynamical pattern is uniquely determined by the phase of the first collision between the beam particles and the scatterer and possesses a universal profile, dictated by the manifolds of the parabolic orbits, which can be understood both qualitatively as well as quantitatively in terms of scattering off a hard wall. We discuss also the inverse problem concerning the possibility to extract properties of the oscillating target from the differential cross sections.
\end{abstract}
 \maketitle

\section{Introduction}
In a scattering process the interaction of the incoming projectiles with the target is of spatially local character, finally yielding free outgoing particles. The main question is to explore the imprints of the interaction potential on the outcoming beam. The basic tool for this study are scattering functions and cross sections relating physical quantities characterizing the outgoing beam with associated quantities of the incoming beam. The scattering angle or the escape (dwell) time as a function of the appropriately defined impact parameter are typical examples of scattering functions while the distribution of the deflection angle of the outgoing particles typically defines the angular cross section. Scattering problems occur in many areas of physics and chemistry \cite{Ott93,Ott02} including celestial mechanics \cite{Hietarinta93,Petit87,Boyd93,Moser01,Benet97,Benet98}, charged particle trajectories in electric and magnetic fields \cite{Chernikov93,Breymann94}, hydrodynamical processes \cite{Novikov79,Aref84,Kedar90}, models of chemical reactions \cite{Rankin71,Noid86,Gaspard89,Koch93} and scattering in atomic, molecular and nuclear physics \cite{Wintgen92,Yuan93}. When the scattering process evolves in two or more spatial dimensions and the interaction
potential is nonlinear, the scattering functions may possess complex structures formed by a fractal set of singularities,originating from homoclinic and heteroclinic intersections of the stable manifolds of unstable periodic orbits (UPOs) occurring in the dynamics of the system. The appearance of such a set of singularities is described by the term {\it chaotic scattering} \cite{Ott93}, an active field of research during the last two decades. In general this set of singularities possesses a non-zero fractal dimension. As shown in Ref. \cite{Jung95} the conclusion on the existence of chaotic scattering via singularities of the scattering functions can be a subtle issue. This is because there exist dynamical systems \cite{Jung95} possessing a countable, infinite, self-similar set of singularities which
do not exhibit topological chaos. On the level of cross sections, which are the most appropriate quantities to be studied in a scattering experiment, the fractal singularities of the scattering functions induce a set of rainbow singularities (RSs) which also possess a fractal geometry \cite{Jung87,Jung05}.

Most of the work done within the framework of chaotic scattering (or scattering in higher dimensions) deals with a static scattering potential. Recently, it has been found that complex behavior with different characteristics can be observed in scattering systems involving a time-dependent scattering potential \cite{Antillon98, Papachristou01,Papachristou04,Papachristou02,Benet05,Benet07,Benet08}. As a representative example of such a system the scattering of freely moving, non-interacting particles off two hard, infinitely heavy, oscillating discs on the plane has been studied. The absence of an energy shell in these systems leads to an increase of phase space dimensionality. The explicit time-dependence of this system leads to a reduction of the impact of the UPOs and their manifolds on the properties of the resulting scattering functions. In Ref.~\cite{Papachristou04} the term dilute chaos was introduced to describe the behavior characterized by the accumulation of peaks, associated with processes leading to maximum energy exchange between scattered particles and an oscillating target around the position of the UPOs of the system. The set of the observed peaks possesses a self-similar structure either in an approximative sense occurring only between two scales determined by the geometry of the particular setup \cite{Papachristou01} or in a local sense in the neighborhood of specific isolated points in phase space \cite{Papachristou04}. The significant role of processes with maximum energy exchange in the scattering off a time-dependent hard potential becomes more transparent in the case of scattering off a single oscillating disc \cite{Papachristou02}. Despite the absence of UPOs in this system the scattering functions possess a nontrivial structure dictated by elementary processes involving single or multiple collisions between the projectile and the oscillating target. Associated with these collision events are processes for which the scattered particle escapes from the interaction region with minimal outgoing velocity. These processes lead to peaks in the scattering function describing the dwell time as a function of the initial velocity of the incoming particle. The term {\it low velocity peaks} (LVPs) was used in Ref.~\cite{Papachristou01} for the description of these structures. In phase space terminology the LVPs are related to the approaching of the trajectories of the scattered particles to the parabolic orbits and their manifolds. The parabolic orbits are the dense set of phase space points lying in the configuration space outside of the interaction region and having velocity equal to zero. The corresponding manifolds consist of phase space points which approach asymptotically the parabolic orbits. A scattering orbit approaches only marginally these manifolds when, due to the energy loss process, it escapes the interaction region with a very low velocity (LV).

So far the analysis of the scattering dynamics in time-dependent scattering potentials has been restricted to the study of scattering functions which, although being a useful methodological tool, are not easily accessible with respect to experimental observation. The aim of this work is to present, for the first time, calculations of differential cross sections, which are observable quantities, for time-dependent scattering off hard potentials on the plane. In this case, in addition to the cross section associated with the distribution of the deflection angle for the outgoing particles, the cross section associated with the distribution of the final velocity of the scattered projectiles is of particular interest, since the energy of the incoming particles is not conserved during the scattering process. It will be shown that when the velocity of the incoming particle is of the same order of magnitude as the maximum velocity of the oscillating scatterer, the corresponding cross sections possess a rich structure consisting of RSs, which can be understood in detail both at a qualitative as well as a quantitative level using the appropriate scattering functions. Our treatment reveals that the key mechanism leading to these singularities is related to the interplay between low-velocity and multiple collision processes. In addition, it will be argued that for a single scatterer of arbitrarily shaped convex boundary the relevant quantity associated with the structures in the cross sections is the phase of the oscillating target at the first impact, which depends on the shape of the scatterer. Within this simple reasoning it is possible also to investigate the inverse scattering problem \cite{Jung99} concerning the extraction of the characteristics of the harmonic movement of a circular scatterer (frequency, amplitude, direction of oscillation) as well as its size, from the profile of the cross sections and scattering functions.

The paper is organized as follows: in Sec. II we present the equations which determine the dynamics of the scattering
problem. We also discuss the role of the various parameters as well as their relevance to the subsequent analysis. In Sec. III we define the scattering functions and cross sections. In Sec. IV we use a simple Gedankenexperiment consisting of a suitably chosen initial beam moving parallel to the oscillation axis of a driven wall, in order to reveal the significance of the phase of the first collision with the moving target for the description and classification of the dynamics in scattering off time-dependent potentials. We introduce a three-dimensional (3D) plot relating the total number of collisions during the scattering process with the initial velocity of the incoming particle and the phase of the oscillating target at the instant of the first collision, as the basic methodological tool allowing the qualitative and quantitative understanding of the scattering dynamics. In Sec. V we make a comparative study of the time-dependent scattering off an oscillating inclined wall with the scattering off an oscillating disk in order to reveal the influence of the geometrical characteristics of the target on the experimentally accessible observables. In Sec. VI we discuss briefly the inverse scattering problem using cross sections and scattering functions to extract dynamical and geometrical characteristics of the scatterer. Finally, we provide in Sec. VII a summary and concluding remarks.

\section{Scattering setup}
\subsection{Scattering dynamics}
\begin{figure}
\includegraphics[width=8.6cm,height=6.45cm]{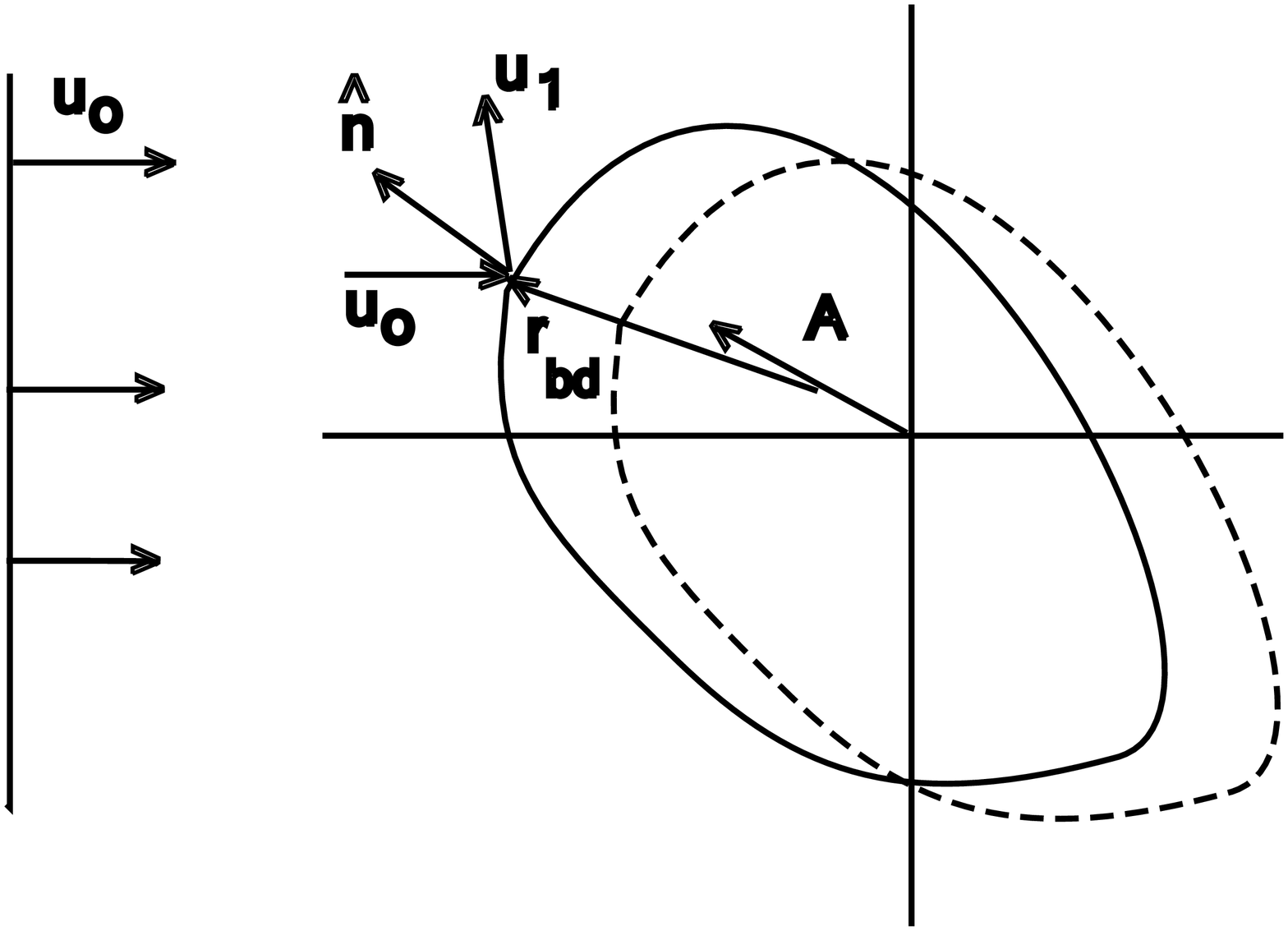}
\caption{An arbitrary scatterer with convex surface is drawn in two positions: when its center is at the coordinate  origin (dashed line) and at a displaced position where an impact takes place (solid line). Indicated is also a beam of particles as well as the important vectors that specify the motion of the scatterer and a projectile before and after the collision event.} \label{fig:Fig1}
\end{figure}

We investigate a dynamical system which, in the general case, consists of a harmonically oscillating scatterer of arbitrary shape with convex boundary in two dimensions, and a beam of particles incident to the interaction region, as shown in Fig.~\ref{fig:Fig1}. The interaction region consists of all the points of the two-dimensional (2D) configuration space, where collisions between the particles and the target are possible. The boundary of the scatterer is impenetrable and its mass is assumed to be much larger than the mass of the particles.
The center of the scatterer, which coincides with the coordinate origin when the scatterer is at its equilibrium position, moves according to the law:
\begin{equation}
\label{eq1} \mathbf{r}_{s}=\mathbf{A}\sin(\omega t+\phi_0)
\end{equation}
where $\omega$ is the frequency, $\mathbf{A}$ is the amplitude vector and $\phi_0$ is the initial phase of the oscillation.

The oscillation angle formed between the positive $x$ semi-axis and the scatterer's oscillation axis is:
\begin{equation}
\label{eq2} \alpha=tan^{-1}\left(\frac{A_y}{A_x}\right)
\end{equation}

The scattering dynamics of a particle with position $\mathbf{r}$ and velocity $\mathbf{u}$ is determined by the equations:
\begin{equation}
\label{eq3} \mathbf{r}_{n+1}=\mathbf{r}_{n}+\mathbf{u}_{n} \Delta t_{n}
\end{equation}
\begin{equation}
\label{eq4}
\mathbf{u}_{n+1}=\mathbf{u}_{n}-2[\hat{\mathbf{n}}\cdot(\mathbf{u}_{n}-\mathbf{u}_{s}(t_{n+1}))]\hat{\mathbf{n}},
\end{equation}
where $n=0,1,2,...,\ell-1$ is the index for the $n$-th collision, $\Delta t_{n}=t_{n+1}-t_{n}$ is the time of free flight in between collisions, $\mathbf{u}_{s}(t)=\mathbf{A}\omega \cos(\omega t+\phi_0)$ is the velocity of the scatterer and $\hat{\mathbf{n}}$ is the normal to the target's boundary at the point of impact. Finally, $\xi_n=\omega t_n +\phi_0$ denotes the phase of the scatterer upon the $n$-th collision. Note that $n=0$ corresponds to the emission time, and thus $\Delta t_{0}=t_{1}-t_{0}$ is the time between emission and first collision. The last collision $\ell$ before the particle escapes from the scattering region, determines the final outgoing velocity $\mathbf{u}_{out}=\mathbf{u}_{\ell}$.

The process where only one collision takes place before the particle leaves the scattering region ($\ell=1$), is the most common one, but there are also processes where more than one collision occurs. In fact we can classify the dynamics of the system into classes of collision events ($\ell$CEs) determined by the total number $\ell$ of collisions before leaving the scattering region.

The equation that determines the time $t_{n+1}$ of the $(n+1)$-st collision is
\begin{equation}
\label{eq5} \mathbf{r}_{n}+\mathbf{u}_{n}\Delta t_{n}=\mathbf{A}\sin(\omega
t_{n+1}+\phi_0)+\mathbf{r}_{bd},
\end{equation}
where $\mathbf{r}_{bd}=(x_{bd},y_{bd})$ is the position vector from the center of the scatterer to the point of impact (see Fig.~\ref{fig:Fig1}). Equation~(\ref{eq5}) is solved numerically, keeping the smallest positive solution.

\subsection{Beam configuration}
The following assumptions specify a typical scattering experiment (see also Sec. V). The beam of identical projectiles is synchronized with the scatterer's oscillation, i.e. the emission time $t_0=0$ is common for all particles and we choose $\phi_0=0$ without loss of generality. The initial position vector is $\mathbf{r}_0=(x_0,b)$, where $x_0$ is the common $x$-coordinate of all particles and $b$ is the impact parameter, which defines the corresponding $y$-coordinate. We use $x_0\ll x_{bd,min}-|A_{x}|$, where $x_{bd,min}$ is the minimal $x$-coordinate of the boundary. We also assume a uniform distribution of $b \in [y_{bd,min}-|A_y|,y_{bd,max}+|A_y|]$, such that the entire projection of the interaction region onto the $y$-axis is covered, as $y_{bd,max}$ and $y_{bd,min}$ are the maximal and minimal $y$-coordinates of the boundary. All particles of the beam initially move parallel to the axis of oscillation, which is chosen to be the $x$-axis ($\alpha=0$), and thus $\mathbf{u}_{0}=(u_0,0)$, $\mathbf{A}=(A_x,0)$, and $b \in [y_{bd,min},y_{bd,max}]$. The
initial magnitude of the particle velocity $u_0$ is the same for all particles of the beam. This choice is compatible with the profile of monochromatic beams often used in scattering experiments since they facilitate the detection of the characteristics of the target.

\subsection{Vectors and parameters of importance}
Three vectors determine the setup of the scatterer and the beam: the initial velocity of the beam particles $\mathbf{u}_{0}$, the amplitude vector of the oscillation $\mathbf{A}$, and the normal to the target's boundary at the point of impact $\hat{\mathbf{n}}$. The first two, have been already chosen parallel to each other lying on the $x$-axis. We will consider in Sec. VI the case of an arbitrary oscillation angle $\alpha\neq 0$, in the context of the inverse problem. The normal $\hat{\mathbf{n}}$ is related to the shape of the target. We intend to address the following cases:
\begin{enumerate}
\item $\hat{\mathbf{n}}=$ constant on the flat surface of an inclined wall (Sec. V A),
\item $\hat{\mathbf{n}}=\hat{\mathbf{n}}(b)$ on the curvilinear boundary of a disk (Sec. V B).
\end{enumerate}

We should bear in mind that in the case of a circular scatterer oscillating parallel to the beam ($\alpha = 0$), $\hat{\mathbf{n}}$ is related to the impact parameter $b$, because each particle of the beam with different $b$ collides with the scatterer at a point with different $\hat{\mathbf{n}}$, which justifies choosing $b$ as an independent variable to define the scattering functions.

What is the main difference between the scattering off static and oscillating targets? When the scatterer is static, the outgoing velocity has always the same magnitude as the ingoing. Only the direction of the outgoing velocity is affected by the shape of the target, and particularly by $\hat{\mathbf{n}}=\hat{\mathbf{n}}(b)$. When the scatterer oscillates, both the magnitude and the direction of the outgoing velocity depend on the target's velocity at the impact time (which in turn is determined by the phase of the oscillation upon collision) as well as on the shape of the scatterer and the impact point on its boundary ($\hat{\mathbf{n}}=\hat{\mathbf{n}}(b)$). Therefore, to examine the effect of the time-dependence of a hard potential, we must consider orbits, which differ with respect to the first collision phase $\xi_1$. We focus on $\xi_1$ and not on the phase of the second or third collision since, as will be shown in the following, $\xi_1$ determines to a large extent the evolution of the orbit and the appearance of $\ell$CEs with $\ell>1$.

Following the above reasoning, it is worth determining the parameters which affect $\xi_1$ in the general case of a parallel beam of particles incident to a hard oscillating target of arbitrary shape (given $\omega$ and A). These are:
(a) the initial velocity, $u_0$, (b) the initial phase of the scatterer when the particle is emitted, $\xi_0=\omega t_0+\phi_0$, (c) the initial position along the $x$-axis, $x_0$ and (d) the initial position along the $y$-axis, i.e. the impact parameter $b$. If one of these quantities changes, keeping the others constant, $\xi_1$ changes too. Therefore, there are four independent ways of affecting $\xi_1$ in time-dependent scattering processes. Our incoming beam is characterized by a single value $u_0$ of the velocities of all particles and varying $u_0$ provides us with the possibility to tune the complexity of the scattering process: Certainly the most interesting behavior has to be expected for velocities $u_0 \approx A\omega$. The remaining three parameters are equivalent to each other. Let us focus on $b$, which is the usually varied parameter in a typical experimental setup (see Sec. II B). For an oscillating scatterer of arbitrary shape, the difference with respect to $\xi_1$ for each particle of the beam originates from the different distance which it has to travel in order to collide for the first time with the scatterer. Therefore, the impact parameter $b$ parameterizes both $\hat{\mathbf{n}}=\hat{\mathbf{n}}(b)$ and $\xi_1=\xi_1(b)$, and thus completely specifies the first collision and its subsequent dynamics.

\section{Differential cross sections and scattering functions}
For the collision experiment we focus on the distributions of (a) the magnitude of the outgoing velocity ${u}_{out}=\sqrt{u_{out,x}^2+u_{out,y}^2}$ and (b) the scattering angle $\theta=tan^{-1}\left(\frac{u_{out,y}}{u_{out,x}}\right)$. These are related to the corresponding differential cross sections:

\begin{equation}
\label{eq6}
\sigma_{q}=\frac{d\sigma}{dq}=\sigma_{t}\cdot\frac{1}{N_{t}} \cdot\frac{dN}{dq}, (q=u_{out},\theta)
\end{equation}
where $N_{t}$  is the total number of particles and $\sigma_{t}$ is the total cross section. Equivalently, one can calculate the respective probability density functions (PDFs): $\varrho_{q}=\frac{1}{N_{t}}\cdot\frac{dN}{dq}$ ($q=u_{out},\theta$), since all the information is included in these quantities while $\sigma_{t}$ is a multiplicative factor.

To explain the properties of the PDFs we will analyze the scattering functions: the outgoing velocity $u_{out}(b)$ and the scattering angle $\theta(b)$ both in terms of the impact parameter $b$. The following relation motivates their usefulness with respect to the analysis of the PDFs:
\begin{equation}
\label{eq7}
\varrho_{q}=\frac{1}{N_{total}}\cdot\frac{dN}{db}\cdot\frac{db}{dq}. (q=u_{out},\theta)
\end{equation}
where $\frac{1}{N_{total}}\frac{dN}{db}=\frac{1}{|y_{bd,min}|+y_{bd,max}}$ is constant.

Of particular importance for our analysis, is the appearance of smooth local extrema of the scattering functions at some value $b=b^*$, such that $\frac{du_{out}}{db}\mid_{b=b^*}=0$ and $\frac{d^2u_{out}}{db^2}\mid_{b=b^*}\neq0$. In this case, the corresponding PDF exhibits a typical square root singularity at $u_{out}(b^*)$ as it can be shown by a Taylor expansion of $u_{out}(b)$ around $b=b^*$ up to order $(b-b^*)^2$
\begin{eqnarray}
\label{eq8}
u_{out}(b)&=&u_{out}(b^*)+\frac{d^2u_{out}}{db^2}\mid_{b=b^*} \cdot(b-b^*)^2.
\end{eqnarray}
If we set $u_{out}=u_{out}(b)$, $u_{out}^*=u_{out}(b^*)$, $\Delta u_{out}=u_{out}-u_{out}^*$ and $\Delta b=b-b^*$, then it follows from Eq.~(\ref{eq8}) that
\begin{equation}
\label{eq9} \frac{\Delta b}{\Delta
u_{out}}\propto\frac{1}{\sqrt{u_{out}-u_{out}^*}}.
\end{equation}
According to Eqs.~(\ref{eq7})-(\ref{eq9}) we conclude that in the
neighborhood of $u_{out}^*$
\begin{equation}
\label{eq10}
\varrho_{u_{out}}\propto\frac{1}{\sqrt{u_{out}-u_{out}^*}}.
\end{equation}

Similarly, smooth extrema of $\theta(b)$ lead to square root singularities in $\varrho(\theta)$. This type of singularities in the cross sections (or equivalently in the PDFs), called rainbow singularities (RSs), have already been observed and analyzed in static scattering processes \cite{Jung05}. In this case a smooth maximum of the scattering function occurs between singularities with fractal structure associated with the intersection of the manifolds of the UPOs of the system. The time-dependent setups considered here do not possess any periodic orbits. Yet, as will be shown in the following, the cross sections possess a set of RSs which can be attributed to parabolic orbits and their manifolds, indicating that the origin of these singularities is of fundamental character (see also the discussion in the introduction). Let us underline here that the positions $b^*$ of the extrema of $u_{out}(b)$ and $\theta(b)$ are identical. This is an important feature that allows to connect certain outgoing velocities with certain scattering angles corresponding to the same RSs as it will be illustrated in the case of the oscillating disk (cf. Sec. V B).

\section{A Gedankenexperiment to understand scattering off time-dependent targets}
\subsection{Gedankenexperiment setup}
In the present section we analyze the structures occurring in the scattering off a vertical hard wall oscillating horizontally according to a harmonic time-law. This setup greatly simplifies the geometry of the scatterer, yet for the usual beam profile of synchronously emitted particles with the same initial velocity ($u_0$) originating from the same location on the $x$-axis ($x_0$), it leads to a single value of $\xi_1$ and consequently to trivial results: Constant scattering functions and delta-shaped cross sections. Since our main interest are properties in the outgoing channel associated with the energy of the scattered particles it seems reasonable to appropriately modify the initial beam, in order to allow for variations of $\xi_1$. To achieve this we introduce an asynchronous particle emission while keeping both $u_0$ and $x_0$ constant or, equivalently, vary $x_0$ while keeping $u_0$ and $\xi_0$ constant. We end up with values of $\xi_1$  covering the interval $[0,2\pi)$ uniformly. However a beam profile containing asynchronously emitted particles  or particles with a certain distribution of the initial distance from the target is not easily at the disposal of the experimentalist and should therefore be considered as a Gedankenexperiment. In the experimental setups with the inclined wall and the disk (see Sec.~V) the interval $[0,2\pi)$ for the values of $\xi_1$ is covered several times corresponding to several intervals of $b$ ($b$-zones). We use the term $b$-zone for the interval of $b$ which corresponds to an interval $[0,2\pi)$ of $\xi_1$. The partitioning of the initial beam into $b$-zones depends on the specific geometry of the scatterer. The main purpose of the Gedankenexperiment is to explicitely illustrate what happens within a single $b$-zone.

An important property of the vertical wall model which simplifies significantly the analysis of the scattering trajectories is that the vector $\hat{\mathbf{n}}$ is constant and always antiparallel to the vector of the incoming velocity. Therefore, Eq.~(\ref{eq4}) becomes:
\begin{equation}
\label{eq11} u_{n+1}=-u_n+2u_s(t_{n+1})
\end{equation}
and the particle leaves the interaction region in opposite direction after a few collisions.

\subsection{Key results of the Gedankenexperiment}

\begin{figure}
\includegraphics[width=8.6 cm,height=13.6 cm]{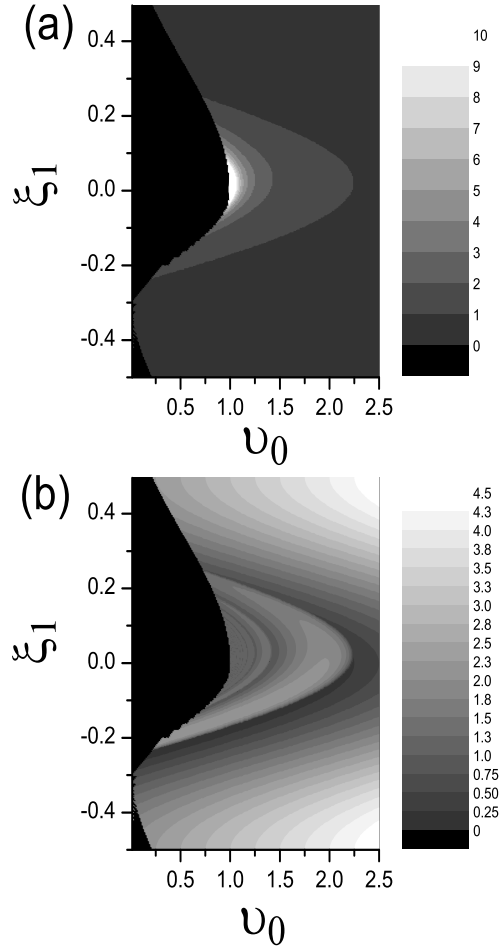}
\caption{3D plots for the vertical wall of the Gedankenexperiment: (a) The total number of collisions $\ell$ as a function of $v_0$ and $\xi_1$, (b) The outgoing velocity $v_{out}$ as a function of $v_0$ and $\xi_1$. Very similar results hold for a single $b$-zone of the inclined wall or the disk.} \label{fig2}
\end{figure}

\begin{figure}
\includegraphics[width=8.6 cm,height=17.2cm]{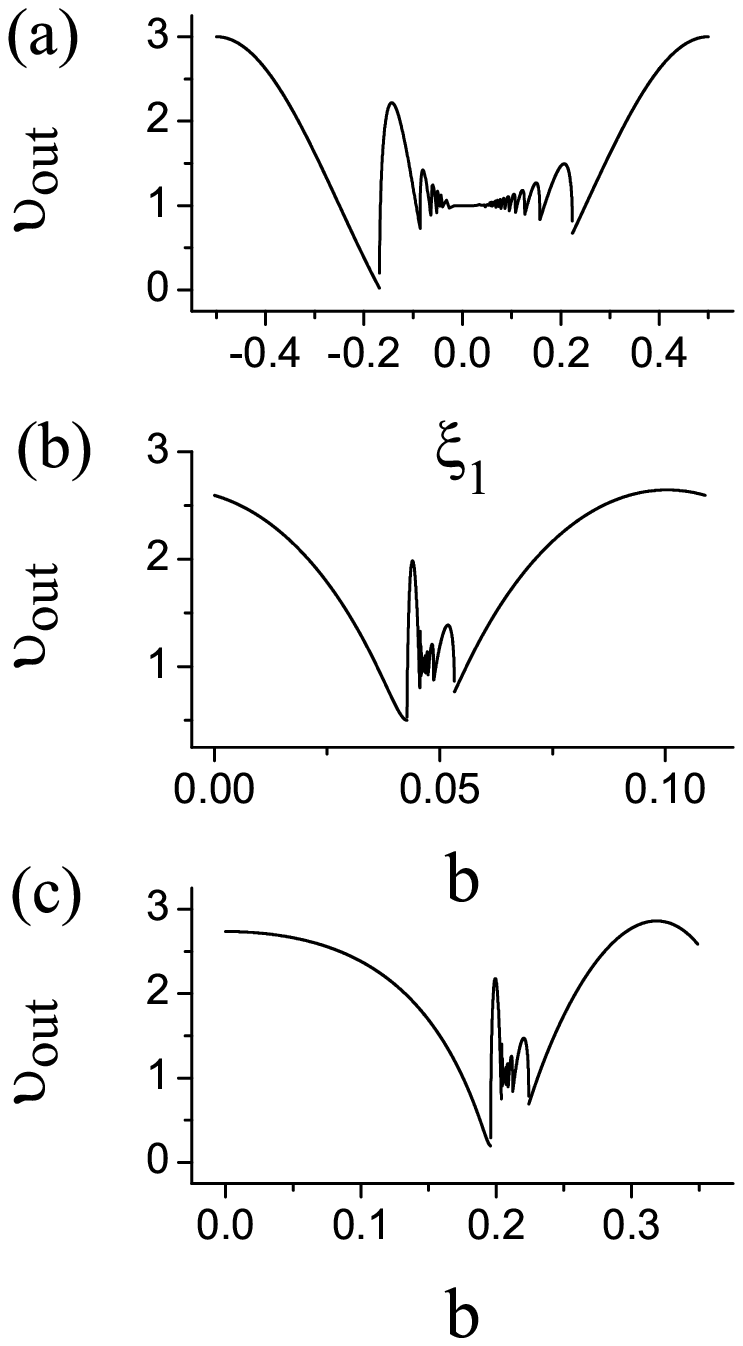}
\caption{(a) The escape velocity $v_{out}$ as a function of $\xi_1$ for $v_0=1$ and the vertical wall. (b) The scattering function $v_{out}$ versus the impact parameter $b$ for a single $b$-zone for the oscillating inclined wall. (c) The same as in (b) but for the oscillating disk.} \label{fig3}
\end{figure}

We employ in the following the dimensionless control parameter $v_0=\frac{u_0}{A \omega}$, that is, the ratio of the initial velocity and the maximum scatterer's velocity. The dynamics of the scattering system is simplified in the limits $v_0 \ll 1$ or $v_0 \gg 1$ where static scatterer approximations are valid. In particular for $v_0 \ll 1$ the dynamics can be approximated by scattering off an effectively static potential obtained by appropriately averaging in time the exact time-dependent potential, while for $v_0 \gg 1$ the movement of the scatterer can be totally neglected and the corresponding observable quantities are similar to those obtained in the case of a static scatterer. Between these two limiting cases the complexity of the dynamics increases and becomes most pronounced with respect to the scattering functions as well as the PDFs for $v_0=1$. We consider the case of small oscillations of the scatterer ($A=0.01$). Lengths are scaled by $A$, velocities are scaled by $A \omega$ (in which case they are denoted by $v$ instead of $u$) and the frequency $\omega$ is set equal to $1$. Finally, $\xi_1$ is scaled by $2 \pi$ (in the figures) with the additional convention that for phases greater than $0.5$ we subtract $1$ in order to get a symmetric representation around $0$.

The influence of the parameter $v_0$ as well as the role of $\xi_1$ is demonstrated in Fig.~\ref{fig2}(a) where the total number of collisions $\ell$ as a function of $v_0$ and $\xi_1$ is plotted. The results of the scattering processes with respect to the outgoing velocity are illustrated in Fig.~\ref{fig2}(b). In Fig.~\ref{fig3}(a) the outgoing velocity $v_{out}$ is presented as a function of $\xi_1 \in [-0.5,0,5)$ for $v_0=1$.

In Fig.~\ref{fig2}(a) we observe that in the region of large $v_0$-values only scattering processes with a single collision event ($1$CEs), as in the static case, occur. As $v_0$ decreases processes where the particle escapes after $2,3,\ldots$ collisions gradually appear. This continues until $v_0 \to 1^+$, where processes with an infinite number of collision events can occur. For $v_0 \leq 1$ the $\ell$CEs with $\ell>1$ disappear one by one rather abruptly, starting from the larger $\ell$ ($\ell \to \infty$), while domains of inaccessible $\xi_1$ appear (black regions). Let us next analyze the associated scattering processes in more detail with respect to $\xi_1$ and $v_0$:
\begin{enumerate}
\item If the first collision takes place when scatterer and incident particle move in opposite directions, (head-on collision), i.e. for $\xi_1 \in [\pi/2,3\pi/2]$ (in the figures $[0.25,0.5]\bigcup[-0.5,-0.25]$) then the particle gains energy and thus escapes after a single collision. This holds for any value of the parameter $v_0$. The maximum outgoing velocity $u_{out,max}$ corresponds to $\xi_1=0.5$ where the magnitude of scatterer velocity is maximum $|-A\omega|$, leading to $u_{out, max}=|-u_{0}-2A\omega|$. Decreasing or increasing $\xi_1$ starting from $0.5$ the particle velocity decreases smoothly (see Fig.~\ref{fig2}(b) and Fig.~\ref{fig3}(a) for $v_0=1$). This simple behavior is responsible for the formation of a smooth maximum of $u_{out}(b)$, characteristic for $1$CE, which in turn implies the appearance of the main RS in the PDF (see Fig.~\ref{fig5}(c)).
\item If the first collision takes place when the scatterer and the incident particle move in the same (positive) direction i.e. $\xi_1 \in [0,\pi/2] \bigcup [3\pi/2,2\pi]$ --or $[-0.25,0.25]$ after
scaling-- then the particle looses energy and more complex processes may occur:
\begin{enumerate}
\item For a large enough value $v_0$ the particle still escapes with a single collision ($\ell=1$). The resulting particle velocity is negative, i.e. the initial velocity of the particle is reversed. The wall velocity is not large enough to change significantly the magnitude of the particle velocity so that the oscillating wall hits the particle once more.
\item When $v_0$ decreases $\ell$CEs with $\ell > 1$ occur as can be seen in Fig.~\ref{fig2}(a). If the first collision takes place near $\xi_1=0$ the wall velocity is large enough to lead to a significant loss of energy for the particle. For 2CEs the particle possesses a negative velocity after the first collision and subsequently the phase of the wall's motion will cover the interval $[0,\pi]$ and finally, as $\xi_2$ approaches $3\pi/2$ from below the second collision occurs (see also Fig.~\ref{fig4}). In the same figure it is also clearly demonstrated that the $2$CE domain increases as $v_0$ decreases. Furthermore, for smaller $v_0$, the velocity of the target near $\xi_1=0$ can make the particle move with a positive velocity after the first collision and thus more than two collisions take place cutting 2CEs' region (for fixed $v_0$) into two pieces. This division into intervals belonging to different values of $\ell$ is also present in Fig.~\ref{fig2}(b). Within each interval we observe that the outgoing velocity $v_{out}$ takes on values from the maximum possible for the corresponding $l$CE to the low velocity (LV) limit. For $v_0=1$ each smooth maxima of $v_{out}(\xi_1)$ in Fig.~\ref{fig3}(a) corresponds to a different $\ell$CE and we observe a global reflection symmetry around $\xi_1=0$. Smooth maxima belonging to neighboring arcs are separated by discontinuous minima (LVPs of the corresponding $\ell$CE).
\end{enumerate}
\end{enumerate}

\subsection{1D representation and analytical calculations}
\begin{figure}
\includegraphics[width=8.6 cm,height=8.45 cm]{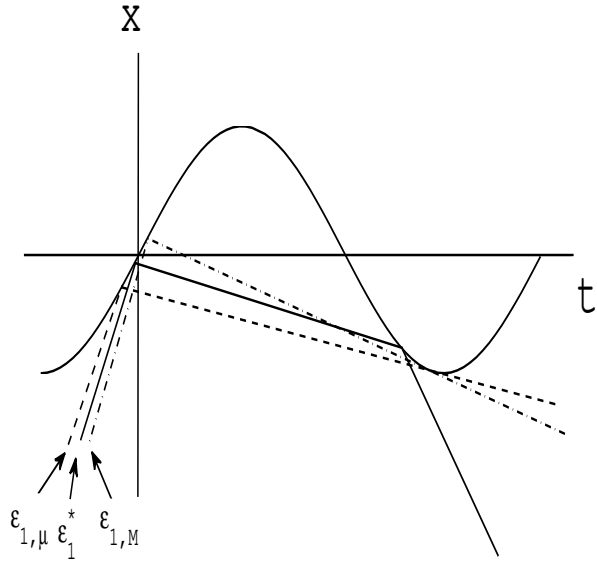}
\caption{1D model of the scattering off an oscillating vertical wall. The sinusoidal curve displays the position of the scatterer as a function of time $t$. The straight line segments represent the movement of different particles which are ejected asynchronously. The LV orbits (corresponding to $\varepsilon_{1,\mu}$ (dashed line) and $\varepsilon_{1,M}$ (dashed-dotted line) near $\xi_1=0$) which define the border between 1CEs and 2CEs are presented. They both approach
tangentially the sinusoidal curve at the second point (near phase $3\pi/2$). The orbit labeled $\varepsilon_{1}^{*}$ (solid line) leads to the maximum outgoing velocity for 2CEs.} \label{fig4}
\end{figure}

An advantage of the simple setup considered in this section is that the dynamics can be represented graphically in an one-dimensional plot since it evolves exclusively along the $x$-axis \cite{Lai91}. A representative plot of this type is shown in Fig.~\ref{fig4} where the position of the wall as a function of time $t$ is drawn as a sinusoidal line. Particle trajectories correspond to a sequence of straight line segments. The beam can be represented in 1D by employing shifts of the initial phase $\xi_0$.

Let us now review the above discussion of the scattering process (see section IV.B) on the basis of Fig.~\ref{fig4}. When $u_0$ is large enough the inclination is also large and, after the reflection of the initial straight line segment, no further intersections with the sinusoidal curve (and consequently collisions with the wall) are possible i.e. we have only 1CEs. Moreover, it is clear that the maximum negative inclination, that is $u_{out,max}$, after the collision appears at $\xi_1=\pi$. Similarly the minimum negative inclination after the collision will appear when $\xi_1=0$.

The above graphical presentation of the dynamics facilitates the derivation of analytical estimates for characteristic quantities determining e.g. the 2CEs region. Decreasing $v_0$ second collisions become possible. If the first collision takes place at $\xi_1=\varepsilon_1$ with $\varepsilon_1 \ll 1$, then the reflected orbit may possess a relatively small negative inclination (velocity) leading to a second collision at $\xi_2=3\pi/2-\varepsilon_2$, $0<\varepsilon_2 \ll 1$. We can approximately determine the corresponding two points $x_1$ and $x_2$, as well as the corresponding velocities of the scatterer $v_{s,1}$ and $v_{s,2}$, by expanding the trigonometric functions determining these quantities in powers of $\varepsilon_i$ ($i=1,2$) (up to terms of order $\varepsilon_i^2$):
\begin{equation}
\label{eq12}
x_1\approx\varepsilon_1,~x_2\approx-1+\frac{{\varepsilon_2}^{2}}{2},~v_{s,1}\approx1-\frac{{\varepsilon_1}^{2}}{2},~v_{s,2}\approx-\varepsilon_2.
\end{equation}
where $x_1$, $x_2$ are scaled by A.

We fix $v_0$ at a typical value allowing for second collisions to take place. Then, we seek for the phases $\varepsilon_1$ such that at the second point ($x_2$) the particle trajectory is tangent to the sinusoidal curve (see Fig.~\ref{fig4}). There are two $\varepsilon_1$ values, denoted as $\varepsilon_{1,\mu}$ and $\varepsilon_{1,M}$, fulfilling this condition. These two phases determine the lower and upper limit which separate the region of 1CE from the the region of 2CEs for a given $v_0$ value. Additionally the particles which trace the border between 1 and 2CEs leave the scattering region with a low velocity (LV) close to the minimum possible for 1CE.The following equations determine the conditions that should be fulfilled for these orbits to be realized:
\begin{subequations}
\label{eq13}
\begin{eqnarray}
x_2-x_1&=&v_1\Delta t_1\omega \\ v_{s,2}=v_1&=&-v_{0}+2v_{s,1}
\end{eqnarray}
\end{subequations}
where $\Delta t_1=(3\pi/2-\varepsilon_2-\varepsilon_1)/\omega$.

Substituting $x_1$, $x_2$, $v_{s,1}$, $v_{s,2}$ in Eqs.~(\ref{eq13}) from Eqs.~(\ref{eq12}), we obtain a 4-th order polynomial equation for $\varepsilon_1$:
\begin{equation}
\label{eq14}
a{\varepsilon_1}^{4}+b{\varepsilon_1}^{3}+c{\varepsilon_1}^{2} +d\varepsilon_1+e=0
\end{equation}
where $a=1/2$, $b=-1$, $c=-3\pi/2+v_0+2$, $d=v_0-1$ and $e=3\pi/2+3+v_0^{2}/2-(3\pi/2+2)v_0$  which can be solved
analytically. We substitute for example $v_0=2.14$, which allows 2CEs, and find two real solutions $\varepsilon_{1,\mu}=-0.1769$ and $\varepsilon_{1,M}=0.4717$, very close to the respective values from
the numerical solution of the system being $-0.1759$ and $0.4709$ respectively (we use here the representation of the phases symmetrically around 0 as done in Fig.~\ref{fig2}(a)). Having found the two values of $\varepsilon_1$, we determine the remaining quantities characterizing the two orbits under study such as the velocities $u_1$, which correspond to 1CEs with minimum outgoing velocity (LVs). Within the interval lying between two successive LVPs a single smooth extremum appears in the scattering function. This property of the LVPs, i.e. their bracketing of smooth extrema of the scattering function, makes them indispensable for understanding scattering processes off time-dependent hard potentials.

It should be emphasized that immediately above the smaller solution $\varepsilon_{1,\mu}$ and just below the greater one $\varepsilon_{1,M}$, the first 2CEs occur. These orbits correspond to the LVs of the 2CEs. Note that the LVs of the 2CEs are somewhat larger than the LVs of 1CEs and thus a discontinuity appears (see Fig.~\ref{fig3}(a)) as the almost tangential intersection of the curve at the second impact point will lead to a more negative inclination of the reflected final line segment, representing the outgoing orbit of the particle.

We next analyze the dynamics in the region $\varepsilon_1 \in (\varepsilon_{1,\mu}$, $\varepsilon_{1,M})$ which corresponds to 2CEs. Starting just above $\varepsilon_{1,\mu}$ and approaching $\varepsilon_{1,M}$, $u_{out}$ initially increases, reaches a maximum (which corresponds to $\varepsilon_1=\varepsilon_1^*$ (solid line in Fig.~\ref{fig4}),
$\varepsilon_{1,\mu}<\varepsilon_1^*<\varepsilon_{1,M}$) and then decreases to the minimum (see Fig.~\ref{fig3}(a)). To determine $\varepsilon_1^*$ we look for the maximum of
\begin{equation}
\label{eq15} v_{out}=v_2=-v_1+2v_{s,2}=-(-v_0+2v_{s,1})+2v_{s,2}
\end{equation}
given that
\begin{equation}
\label{eq16} x_2-x_1=v_1\Delta t_1\omega,
\end{equation}
where $\Delta t_1=(3\pi/2-\varepsilon_2-\varepsilon_1)/\omega$. In other words, we look for $\varepsilon_1^*$, such that:
\begin{equation}
\label{eq17}
\frac{dv_{out}}{d\varepsilon_1}\mid_{\varepsilon_1=\varepsilon_1^*}=0.
\end{equation}

We substitute $x_1$, $x_2$, $v_{s,1}$, $v_{s,2}$ from Eqs.~(\ref{eq12}) into Eqs.~(\ref{eq15}-\ref{eq16}) to obtain a
system of 2 equations with 3 variables $\varepsilon_1$, $\varepsilon_2$, and $v_{out}$. Subsequently we find the relation $v_{out}=f(\varepsilon_1)$ after the substitution of $\varepsilon_2$. Applying the condition (\ref{eq17}) yields $\varepsilon_1^*=0.1472$ for $v_0=2.14$, which is close to the numerical value $0.1320$. This smooth maximum of 2CEs, i.e $u_2$ which corresponds to $\xi_1=\varepsilon_1^*$, causes the secondary RSs in the PDFs. We remind the reader of the fact that the main smooth maxima that appear in the scattering functions (and correspond to the main RSs) in the PDFs are those of the 1CEs, because they always cover a greater width with respect to $\xi_1$ as they always include half of the complete interval of $\xi_1$, namely $[\pi/2,3\pi/2]$. Additionally, if we are interested in the secondary RSs we should study all smooth maxima of $\ell$CEs with $\ell >1$ which appear, as it will be shown, close to those of the 2CEs.

The last calculation concerns the critical $v_0^{cr}$ which is the threshold for 2CEs (see Fig.~\ref{fig2}(a)). Note that $v_0$ is a parameter in Eq.~(\ref{eq14}) for $\varepsilon_1$. 2CEs emerge first when the two real solutions $\varepsilon_{1,\mu}$ and $\varepsilon_{1,M}$ are equal (see Fig~\ref{fig4}). Therefore we ask for the value of the parameter $v_0$ ($v_0^{cr}$), such that we encounter one double real solution ($\varepsilon_1^{cr}$) of Eq.~(\ref{eq14}). We finally find $v_0^{cr}=2.237$ which lies very close to the value $2.235$ obtained numerically.

Using similar arguments we can perform the corresponding calculations for the remainder of the $\ell$CEs ($\ell>2$), which appear gradually as $v_0$ decreases. The limits of the region of 3CEs for a certain $v_0$ can be approximated if we suppose that the orbits we seek have $\xi_1$ near 0, second collision with a phase close to $\pi/2$ and a third collision tangential to the curve at the point with phase $\pi$. Note that the LVs of the 3CEs are slightly greater than those of the 2CEs, whereas their corresponding maxima are a bit lower in close analogy with the comparison of 1CEs and 2CEs. If we proceed with decreasing further $v_0$, we should observe the appearance of 4CEs that will divide the area of 3CEs into two pieces. For the approximation of the limit orbits for $\ell$CEs with $\ell>3$ the same two last conditions with $3$CEs should be satisfied, but one should further demand that more collisions occur in the phase interval [$0,\pi/2$]. This process continues until $v_0=1$ where we can have an infinite number of collisions, i.e. the particle follows the orbit of the scatterer.

For $v_0<1$ the crucial difference is that $u_0$ is smaller than the maximum velocity $A \omega$ of the scatterer, i.e. some values of $\xi_1$ (especially those near $\xi_1=0$) are no more accessible (see Fig.~\ref{fig2}(a)). This fact causes an abrupt disappearance of $\ell$CEs with $\ell>1$, starting from larger $\ell$ and proceeding to the smaller ones, so that finally the area of 1CEs shrinks until it becomes vanishingly small as $v_0 \to 0$.

Lets conclude our Gedankenexperiment. The value of the parameter $v_0$ which corresponds to the richest behavior of the scattering processes is $v_0=1$. Several $\ell$CEs appear, separated from the respective LVPs, with smooth maxima in between (see Fig.~\ref{fig3}(a)), which in turn correspond to RSs in the PDF $\varrho_{u_{out}}$ (see Fig.~\ref{fig5}(c) for example). Given that LVs are defined as the lowest possible outgoing velocities for an $\ell$CE scattering process, they constitute the closest trajectories to the parabolic orbits and their manifolds. This property of the LVPs allows us to connect the RSs observed in the PDF with the parabolic orbits as well as their manifolds in phase space.

\section{The influence of the geometry on the scattering dynamics}
\subsection{Oscillating inclined wall}
The inclined wall has finite size (this choice simulates the finite size effect in the case of an oscillating disk) such that the scattering region in the y-direction is entirely covered by a beam with $b \in [-1,1]$. We examine here the influence of the inclination of the wall $\gamma$ i.e the acute angle between
the wall and the positive horizontal semi-axis $x$. For all particles the normal is the same $\hat{\mathbf{n}}=(-\sin\gamma,\cos\gamma)$ and the unitary tangential vector is also constant
$\hat{\mathbf{t}}=(\sin\gamma,\cos\gamma)$. In the case of a static inclined wall we have simply $u_{out}=u_{0}$
and $\theta=2\gamma$, i.e. the scattering functions are constant and PDFs delta function like.

\begin{figure}
\includegraphics[width=8.6cm,height=17.2cm]{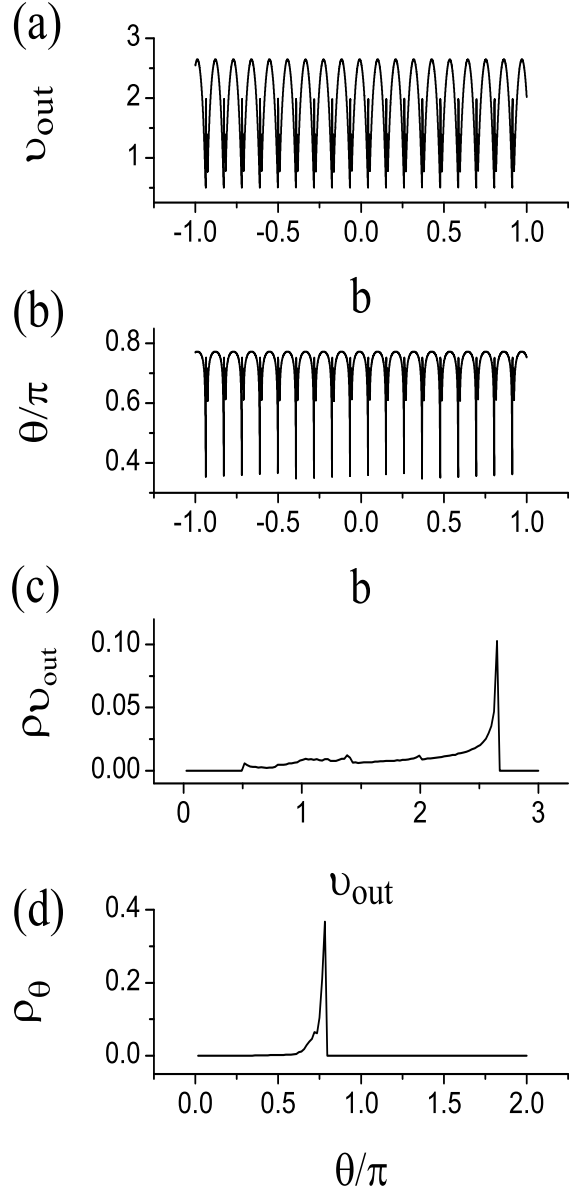}
\caption{Characteristic quantities for the scattering off the oscillating inclined wall ($\gamma=\pi/3$, $v_0=1$): The scattering functions (a) $v_{out}(b)$, (b) $\theta(b)$ and the PDFs (c) $\varrho_{v_{out}}$, (d) $\varrho_{\theta}$.} \label{fig5}
\end{figure}

In the case of an oscillating inclined wall ($\gamma=\pi/3$) there is a continuous variation of $\xi_1$ with the impact parameter $b$. This originates from the difference with respect to the distance which the particles have to travel due to the inclination of the wall. Therefore, it is expected that the structure found in the Gedankenexperiment (see Fig.~\ref{fig3}(a)), discussed in the previous section, appears here too. Indeed, in the scattering function of the outgoing velocity $v_{out}(b)$ (see Fig.~\ref{fig5}(a)) we observe a repetition of the structures belonging to a single b-zone. This is due to the fact that $\xi_1$ is not restricted to the range of one cycle $[0,2\pi)$ (as in the Gedankenexperiment), but, depending on the inclination and the length of the wall, the particles can collide during several cycles of the oscillation. We can divide the impact parameter in $b$-zones each of which corresponds to values of $\xi_1$ covering -continuously but not uniformely- a period of oscillation. The number of the repetitions of the structure is equal to the number of the $b$-zones, $N_{zn}=\frac{2}{tan(\gamma)u_{0}\frac{2\pi}{\omega}}$. A typical characteristic of the inclined wall is that $b$-zones have equal length and for this reason we observe a pattern consisting of a repetition of a single structural element. However, this is not an exact repetition, because the finite size of the scatterer introduces a slight difference between the boundary zones and the internal ones. The total number of collisions $\ell$ as a function of $v_0$ and $\xi_1$ for a $b$-zone of the inclined wall ($b \in [0,tan(\gamma)u_{0}\frac{2\pi}{\omega}]$ is very similar to the corresponding one for the vertical wall (see Fig.~\ref{fig2}(a)). Consequently, for a single $b$-zone and for $v_0=1$, we expect a similar structure in the scattering function $v_{out}(b)$ as that of Fig.~\ref{fig3}(a) of the Gedankenexperiment and this is indeed the case as we can see in Fig.~\ref{fig3}(b).

The difference between the direction of the normal $\mathbf{\hat{\mathbf{n}}}$ and the initial velocity $\mathbf{u}_0$, affects scattering angles, and thus $\theta\neq \pi$ in contrast to the case of the vertical wall. The scattering function $\theta(b)$ is shown in Fig.~\ref{fig5}(b) where we can observe the $b$-zone pattern discussed above.

To ease the interpretation of the result of the scattering process we can rewrite Eq.~(\ref{eq4}) in terms of the normal and tangential components:
\begin{equation}
\label{eq18}
\mathbf{u}_{n+1}=(\mathbf{u}_{n}\cdot\mathbf{\hat{t}})\mathbf{\hat{t}}-
[\hat{\mathbf{n}}\cdot(\mathbf{u}_{n}-2\mathbf{u}_s(t_{n+1}))]\hat{\mathbf{n}}
\end{equation}

In case we are interested only in 1CE we can put $n=0$ and $u_{1}=u_{out}$. Obviously, the tangential component does not change whereas the normal one depends on the velocity of the wall at the impact time ($\xi_1$). When moving initially with a velocity of opposite sign compared to that of the wall i.e. $\xi_1 \in (\pi/2,3\pi/2)$ the particle gains energy and approaches $\hat{\mathbf{n}}$, so it has a larger outgoing velocity and scattering angle. Therefore, $\theta(b)$ and $v_{out}(b)$ possess a similar appearance. The maximum of $v_{out}$ and $\theta$ corresponds to $\xi_1=\pi$, where the wall velocity is minimum.
\begin{equation}
\label{eq19} v_{out,max}=\sqrt{2+v_0(2+v_0)-2(1+v_0)\cos 2 \gamma}
\end{equation}
\begin{equation}
\label{eq20} \theta_{max}=tan^{-1}\left(\frac{(1+v_0)\sin 2
\gamma}{-1+(1+v_0)\cos 2 \gamma}\right)\leq\pi/2+\gamma
\end{equation}

In Figs.~\ref{fig5}(c) and \ref{fig5}(d) the PDFs $\varrho_{v_{out}}$ and $\varrho_{\theta}$ are presented,
respectively. We can verify that the main RS of the PDFs corresponds to the smooth maxima for 1CE derived above. For
$\gamma=\pi/3$ we obtain from Eqs.~(\ref{eq19})-(\ref{eq20}) $v_{out,max}=2.65$ and $\theta_{max}/\pi=0.77$  which coincide with the numerical results. It should be noted that for all the $b$-zones in the scattering functions these smooth maxima are the same, so they will contribute to the same main peak in the cross sections increasing its height.

Secondary RSs correspond to the maxima of $\ell$CEs with $\ell >1$. In $\varrho_{v_{out}}$ we observe secondary RSs close to $v_{out}=2.0$ (corresponding to 2CEs maximum), close to 1.35 (3CEs maximum), and an accumulation of overlapping maxima for $v_{out} \in (0.7,1.3)$ but still leading to very minor peaks ($\ell$CEs maximum with $\ell >3$). The remote peak at $v_{out}=0.5$ corresponds to the LVP for 1CE. Secondary RSs for $\varrho_{\theta}$ are very close to the dominant 1CE peak (as we can observe in the corresponding scattering function $\theta(b)$) and thus contribute to the increase of the width of this peak.

\subsection{Oscillating disk}

Let us now examine the influence of the curvature on the structure of scattering functions and consequently on the PDFs by investigating the scattering off an oscillating disk (radius $R=1$ and center placed at the origin). The normal unit vector $\hat{\mathbf{n}}$ at each point of the scatterer's boundary depends for $\alpha=0$ only on the impact parameter $b$:
\begin{equation}
\label{eq21}
\hat{\mathbf{n}}=(-\sqrt{1-\frac{b^2}{R^2}},\frac{b}{R}).
\end{equation}

In addition the unit tangent vector reads:
\begin{equation}
\label{eq22}
\hat{\mathbf{t}}=(-\frac{b}{R},-\sqrt{1-\frac{b^2}{R^2}}).
\end{equation}

In the case of a static disk the magnitude of the outgoing velocity is equal to the initial one ($v_{out}(b)=u_{0}$ and
$\varrho_{u_{out}}=\delta(u_{out}-u_{0})$) but the scattering angle varies smoothly because of the curvilinear boundary of the disk (variation of $\hat{\mathbf{n}}$):
\begin{equation}
\label{eq23} \theta(b)=\tan^{-1}\left(\frac {2b\sqrt{R^2-b^2}}{2b^2 - R^2} \right)
\end{equation}
\begin{equation}
\label{eq24}
\varrho_{\theta}=\frac{|\sin\left(\theta/2\right)|}{4}
\end{equation}
This form of $\theta(b)$  remains as a backbone in the case of an oscillating disk.

\begin{figure}
\includegraphics[width=8.6cm,height=17.2cm]{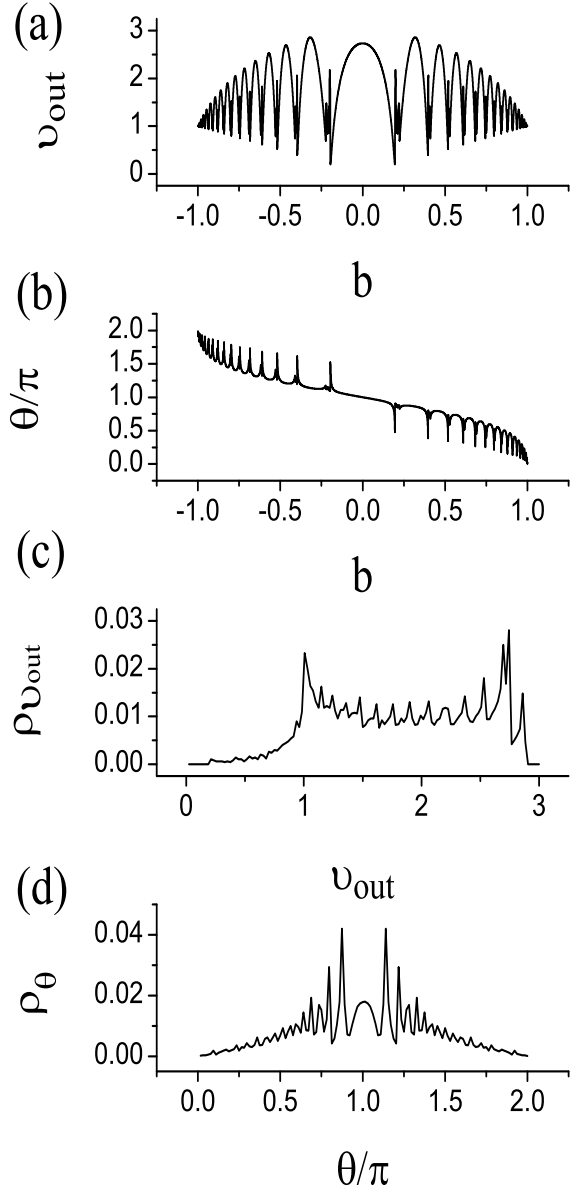}
\caption{Characteristic quantities for the scattering off the oscillating disk ($\alpha=0$, $v_0=1$): The scattering functions (a) $v_{out}(b)$, (b) $\theta(b)$ and the PDFs (c) $\varrho_{v_{out}}$, (d) $\varrho_{\theta}$.} \label{fig6}
\end{figure}

When the disk oscillates, the variation of $\xi_1$, arises due to the different distances that the particles have to travel because of the curvature. Thus, we expect a repetition of the basic structure of the $b$-zone of the Gedankenexperiment to occur in the scattering function $v_{out}(b)$ (see Fig.~\ref{fig6}(a)). In Fig.~\ref{fig3}(c) the scattering function $v_{out}(b)$ for a $b$-zone is presented, exhibiting a structure similar to that observed in Fig.~\ref{fig3}(a) for the Gedankenexperiment. Nevertheless, each particle collides at a different point on the scatterer with a different normal and tangential vector and not a constant one as is the case for the wall. This fact influences the values of $v_{out}$ for each $b$-zone, as well as the partitioning into $b$-zones, which are not anymore of the same length, but become narrower as we approach $b=R$ or $b=-R$. Namely, the upper limit $b_z$ of the $z$-th zone ($z=1,2,..,N_{zn}=\frac{2R}{u_{0}\frac{2\pi}{\omega}}$)  is:

\begin{equation}
\label{eq25}
b_z=\sqrt{zu_{0}\frac{2\pi}{\omega}(2R-zu_{0}\frac{2\pi}{\omega})}
\end{equation}

Using Eq.~(\ref{eq25}) we can calculate the length of each zone: $b_z-b_{z-1}$ with $b_0=0$. For the zones $b<0$ the same relation (r.h.s of Eq.~(\ref{eq25} multiplied by $-1$) holds for the lower limit of the zone. Equation (\ref{eq25}) also describes the accumulation of the maxima towards the boundaries of $b$ ($\to \pm R$).

In Fig.~\ref{fig6}(b) the scattering function $\theta(b)$ is presented. One can observe smooth minima and maxima for $b < 0$  and $b > 0$, respectively. This happens because $\hat{\mathbf{n}}$ points to different directions below and above $b=0$, namely towards the scattering angles $(\pi,3\pi/2)$ and $(\pi/2,\pi)$, respectively. The latter affects the scattering angle of a particle that (for example) approaches $\hat{\mathbf{n}}$ (gains energy) in the following way: for $b>0$ the scattering angle becomes larger as the trajectory after the collision approaches $\hat{\mathbf{n}}$, whereas for $b<0$ it becomes smaller. The inverse happens when the particle diverges from $\hat{\mathbf{n}}$ because of energy loss.

We stress once more that there is a one-to-one correspondence of the structures (location of extrema and LVPs) appearing in $\theta(b)$ and $v_{out}(b)$. Therefore the RSs of the PDFs  $\varrho_{v_{out}}$ and $\varrho_{\theta}$ (Fig.~\ref{fig6}(c) and \ref{fig6}(d), respectively) originating from the same extrema of the corresponding scattering functions are uniquely related to each other. Another important characteristic of the PDFs $\varrho_{v_{out}}$ and $\varrho_{\theta}$ in the case of the oscillating disk, is that the number of the main RSs which correspond to the
smooth extremum of each $b$-zone for $1$CEs, are equal to the number of the b-zones. This discrimination of RSs, which did not exist for the inclined wall, originates from the curvature (change of $\hat{\mathbf{n}}$) which forms unequal values of the extrema in the scattering functions for each $b$-zone (see Fig.~\ref{fig6}(a)). Besides the extrema originating from $1$CEs, which correspond to the main RS, there occur extrema within each $b$-zone in the scattering functions which stem from $\ell$CEs with $\ell>1$. These extrema have different but very close values in each $b$-zone (see Fig.~\ref{fig6}(a)), and thus contribute to the wide peak of $\varrho_{v_{out}}$ at approximately $v_{out}=1.0$ (see Fig.~\ref{fig6}(c)). The fluctuations in this diagram for $0<v_{out}<0.8$ stem from events with minimum outgoing velocity (LVPs). We can verify this by considering the corresponding scattering function $v_{out}(b)$ shown in Fig.~\ref{fig6}(a). The analysis of $\varrho_{\theta}$ (Fig.~\ref{fig6}(d)) is similar to the preceding one for $\varrho_{u_{out}}$. We observe RSs corresponding to the maxima and minima of the scattering function with a symmetry
around $\theta=\pi$. Each peak of this diagram corresponds to a peak of $\varrho_{v_{out}}$ and thus in the case of an oscillating disk, we can redistribute the particles of a plane wave like beam to certain angles and associated velocities within the scattering process.

An alternative way of interpreting the scattering functions and PDFs of the oscillating disk is presented in
the Appendix. If we are interested only in the main peaks of the PDFs resulting from $1$CEs, then we can use the
approximation that the disk does not move in configuration space but only changes velocity according to the law
$\mathbf{u}_{d}=\mathbf{A}\omega \cos(\omega t+\phi_0)$ (static disk approximation SDA)\cite{Lichtenberg:1992,Lieberman:1972}. We derive the following analytical expressions for $u_{out}$
\begin{eqnarray}
\label{eq26} u_{out}=&(1/R)(R^2 u_0^2 + 2 A \omega \cos(\zeta)\times
\nonumber\\
&(2(b^2-R^2) u_0 \cos(\alpha) + 2 b \sqrt{R^2-b^2} u_0 \sin(\alpha)
+ A \omega \cos(\zeta)\times
\nonumber\\
&(R^2+(R^2-2b^2)\cos(2\alpha)-2b\sqrt{R^2-b^2}\sin(2\alpha))))^{1/2},
\end{eqnarray}
where $\zeta=\frac{\phi_0 u_0 - (\sqrt{R^2-b^2}+x_0) \omega}{u_0}$, and $\theta$
\begin{equation}
\label{eq27}
  \theta_1(b)=tan^{-1}\left( \frac
  { 2b [u_{0} \sqrt{R^2-b^2} + A \omega \cos(\zeta) (b \sin(\alpha) - \cos(\alpha) \sqrt{R^2-b^2} ) ] }
 { (2b^2-R^2) u_{0} - 2 A \omega \cos(\zeta) (b \sin(\alpha) \sqrt{R^2-b^2}+ \cos(\alpha)(b^2-R^2))  }
\right)
\end{equation}
and
\begin{equation}
\label{eq28}
 \theta_2(b)=tan^{-1}\left( \frac
{ 2 A \omega b \cos(\zeta) (\cos(\alpha)\sqrt{R^2-b^2} - b
\sin(\alpha)) } { R^2u_{0}+ 2 A \omega \cos(\zeta) \cos(\alpha)
(b^2-R^2) + 2 A \omega b \cos(\zeta) \sin(\alpha) \sqrt{R^2-b^2}  }
\right).
\end{equation}
for $\mathbf{g}\geq0$ and $\mathbf{g}<0$ respectively, $\mathbf{g}=(2\mathbf{u}_d-\mathbf{u}_{0})\cdot\hat{\mathbf{n}}$.
Please note that $\cos(\zeta)$ appears both in $u_{out}(b)$ and $\theta(b)$. If we are interested in the values of the impact parameter $b=b^*$ where an extremum of the scattering function is located, we require $\frac{d\cos(\zeta)}{db} \mid_{b=b^*}=0$ that is:
\begin{equation}
\label{eq29} \frac{ b^* \omega \sin(\zeta)} {u_0 \sqrt{R^2-b^{*2}}}=0
\end{equation}
An obvious solution is $b^*=0$. The other solutions can be derived
from the condition $\zeta=m \pi$ which gives:
\begin{equation}
\label{eq30} b^*=\pm \sqrt{R^2-x_0^2-\frac{m^2 \pi^2
u_0^2}{\omega^2} + \frac{2 m \pi u_0^2 \phi_0}{\omega^2} -
\frac{u_0^2 \phi_0^2}{\omega^2} + \frac{2 m \pi u_0 x_0}{\omega} -
\frac{2 u_0 x_0 \phi_0}{\omega}}
\end{equation}
where $m$ is an integer between $\frac{u_0 \phi_0 - R \omega + x_0 \omega}{\pi u_0}$ and  $\frac{u_0 \phi_0 + R \omega +x_0 \omega}{\pi u_0}$.

\begin{figure}
\includegraphics[width=9cm,height=8.4cm]{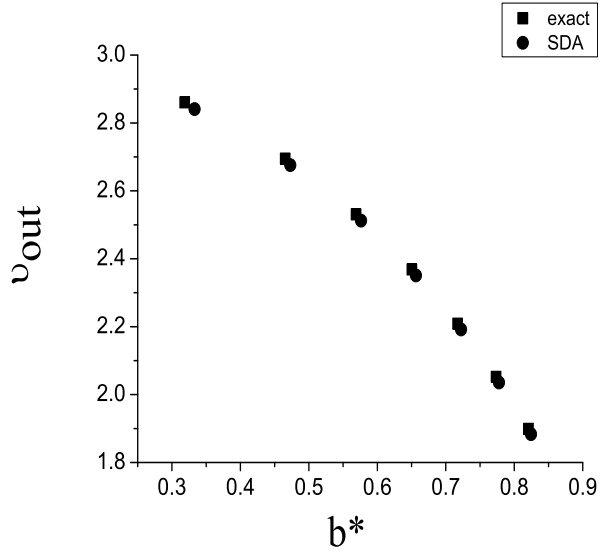}
\caption{The values of successive extrema of $v_{out}$ corresponding to $b^*$ obtained utilizing the exact model ($\square$) and the SDA ($\bigcirc$).} \label{fig7}
\end{figure}

In Fig.~\ref{fig7} we present the values given by the exact numerical solution and the SDA for $b^*$ and $v_{out}(b^*)$. We observe an overall good agreement. These values of $b^*$ and $v_{out}(b^*)$ are only approximately independent of $\xi_0$. However, it should be stressed that for values of $u_{out}(b)$ close to $b=0$ the SDA ceases to be a good approximation. This is because the effect on $v_{out}$ of the disk's motion in configuration space (which is neglected by SDA) becomes comparable with the effect of the curvature. For the same reason the dependence on $\xi_0$ becomes very important for these values. On the contrary, near $b \to R $ the curvature forms to a large extent the outgoing velocity of the particle and SDA predicts accurately $u_{out}$, which is also approximately independent of $\xi_0$.

\section{The inverse scattering problem}
Here we assume that the shape of the hard scatterer is circular and the oscillation law is harmonic. An experimentalist could estimate $\omega$ using a pulsed beam with certain repetitions in time. If the corresponding time intervals become equal to the period of the oscillation the synchronization will result in discriminating the peaks of the PDFs that otherwise should have been smoothened out due to the phase averaging. The oscillation amplitude $A$ can be obtained from the maximum outgoing velocity observed, which always obeys the equation $u_{out,max}=u_0+2A\omega$. The radius of the scatterer $R$ is related to the number of dominant peaks in the PDF for $v_{out}>1$ (see Fig.~\ref{fig6}(c),(d)) which is equal to the number of $b$-zones provided by the equation $N_{zn}=\frac{2R}{u_{0}\frac{2\pi}{\omega}}$.

\begin{figure}
\includegraphics[width=8.6cm,height=17.2cm]{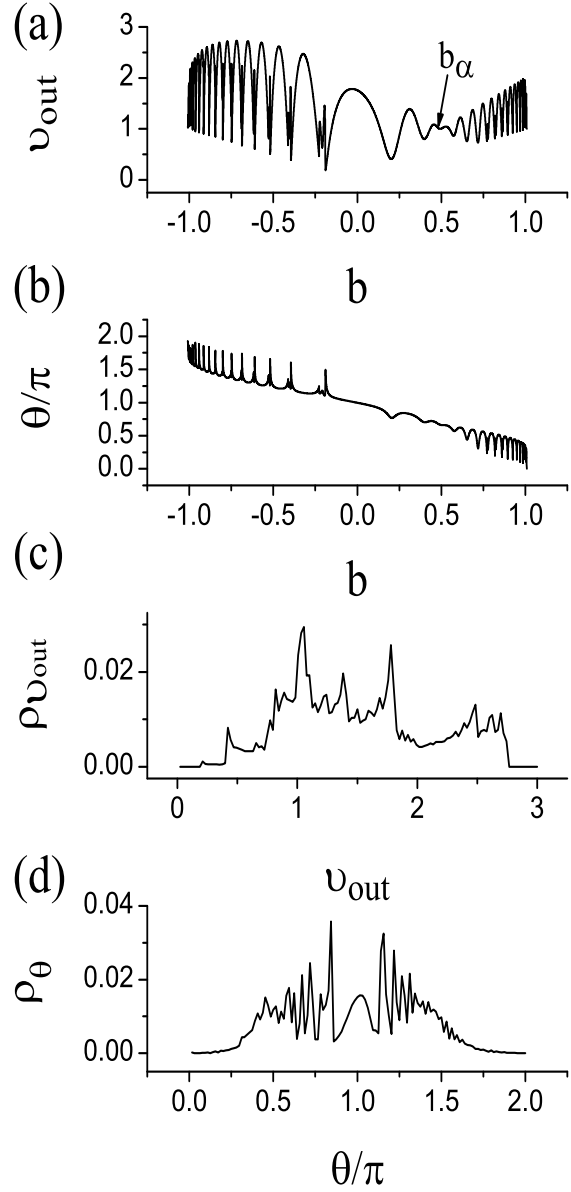}
\caption{Characteristic quantities for the scattering off the disk with
oscillation angle $\alpha=\pi/3$: The scattering functions (a) $v_{out}(b)$, (b) $\theta(b)$ and the PDFs (c) $\varrho_{v_{out}}$, (d) $\varrho_{\theta}$.} \label{fig8}
\end{figure}

As soon as $\omega$, $A$, and $R$ have been obtained there is only one information missing to complete the description of the disk's motion, namely the oscillation angle $\alpha$. If we encounter a nonzero angle $\alpha$ we observe a breaking of the reflection symmetry around $b=0$ in the scattering function $v_{out}(b)$ (see Fig.~\ref{fig8}(a) for $\alpha=\pi/3$) with corresponding results for the PDF $\varrho_{v_{out}}$ (Fig.~\ref{fig8}(c)). We can quantify this asymmetry by the position $b_{\alpha}$ of the point which is determined via the condition $\mathbf{u_s} \cdot \hat{\mathbf{n}}=0$ leading to zero momentum change of the projectile due to the motion of the disk (see Eq.~\ref{eq4}). This yields:
\begin{equation}
\label{eq31} b_{\alpha}=R \cos\alpha.
\end{equation}
For $\alpha=0$, $b_0=R$ corresponds to the accumulation of the RSs (see Fig.~\ref{fig6}(a)). For $\alpha \ne 0$, $b_{\alpha}$ provides us with the position of the minimum of the upper envelope of $v_{out}(b)$ moving closer to $b=0$ with increasing $\alpha$ (see Fig.~\ref{fig8}(a)). We stress that the variation of the normal vector $\hat{\mathbf{n}}$ with varying $b$ is greater in the vicinity of $b=R$ compared to the neighborhood of $b=0$. As a result the shift of the accumulation point $b_{\alpha}$ towards $b=0$ affects the overall appearance of the scattering function which exhibits smooth oscillations near this point. This is because near the point on the surface corresponding to $b_{\alpha}$ the effective velocity of the disk, i.e. $\hat{\mathbf{n}}\cdot\mathbf{u}_{s,n+1}$ in Eq.~(\ref{eq4}) is very small and the dynamics approach the scattering from a static disk. Consequently the region of smooth behavior of e.g. $v_{out}(b)$ near $b_{\alpha}$ is shifted closer to $b=0$ and its size increases.

Regarding the PDF the variation of the oscillation axis of the scatterer, does lead to changes of both the width and the location of the RSs. However, the location of these peaks also depends on the initial phase of oscillation $\xi_0$. This behavior becomes even more evident as $b^*$ approaches 0 (see Sec. V B). Moreover a phase averaging leads to a smooth PDF without dominant peaks. Therefore, the location of the RSs with respect to $\varrho_{v_{out}}$ does not possess a one-to-one correspondence with the oscillation angle, and consequently cannot be used for determine it, at least for the case of an unknown $\xi_0$.

The effect of the rotation of the oscillation axis is shown in Fig.~\ref{fig8}(b) for $\theta(b)$ where we observe a shift of the position $b_{\alpha}$ of the minimum of the upper envelope, and in Fig.\ref{fig8}(d) for the PDF where a breaking of the symmetry with respect to $\theta=\pi$ occurs.

\section{Concluding remarks}
In the present work, we have investigated the scattering of a beam of non-interacting particles off an oscillating target in the plane. Our study focuses on the exploration of the basic scattering mechanisms and on the experimentally accessible differential cross sections for the outgoing velocity of the scattered particles and the deflection angle. At the methodological level we have elevated the importance of the phase of the oscillating target at the instant of the first collision for the description and classification of the collision events comprising the scattering process. In order to gain additional insight into the scattering dynamics we have considered the simplest possible system consisting of an oscillating wall and a beam of particles moving with a constant velocity parallel to the axis of oscillation of the wall. We have shown that within this simple example the complete complexity of the scattering dynamics is recovered in a Gedankenexperiment with asynchronous ejection of the particles in a beam. In this system the ensemble of the particle trajectories is divided in subsets each containing orbits with a fixed number of total collisions between particle and oscillating wall. This classification has been clearly illustrated in a 3D plot displaying the total number of collisions $\ell$ between particle and wall as a function of the initial velocity $u_0$ of the particle and the phase of the scatterer oscillation at the instant of the first collision $\xi_1$. A closer investigation of the underlying dynamics reveals a universal behavior expressed through specific critical points separating regions in the $(u_0,\xi_1)$ plane formed by trajectories with different values of $\ell$. This behavior turns out to be generic for scattering off oscillating targets as suggested by our analysis of the scattering dynamics for two other examples, the inclined wall and the disk. In addition we have shown that each zone with a fixed number of collisions and $\xi_1 \in [0,2 \pi]$ is characterized by a smooth maximum in the scattering functions relating the impact parameter of the incoming beam of particles with the outgoing velocity or the deflection angle. Each smooth maximum leads to a RS expressed through a prominent peak with prescribed width in the associated differential cross section (or equivalently the PDF). The position of the peak in the cross section is given by the phase space location of the maximum while its width is determined through the locations of the events leading to minimum outgoing velocity, bracketing the smooth maximum, in the related scattering function. It is the latter property which allows us to connect the peaks occurring in the cross section of the considered scattering processes with the parabolic orbits and their manifolds in the corresponding phase space. In fact the region of multiple collisions is accessible when the velocity of the beam particles becomes comparable with the maximum velocity of the scatterer. In this case the cross sections of the scattering process attain a characteristic profile, generated by the superposition of peaks attributed to RSs. The complexity of the observed pattern increases as the velocity of the particles approaches the maximum velocity of the oscillating scatterer due to the accumulation of the peaks in this limit. It is a unique characteristic of our time-dependent system  that the RSs in the cross section of the outgoing velocity possess a one-to-one correspondence with those of the scattering angle. It is important to notice here that both the RSs in the cross section, as well as the singularities in the scattering functions form a finite set. No topological chaos is present in the scattering off the harmonically oscillating disk.

Our analysis is valid for hard scattering off an arbitrarily shaped oscillating target with convex geometry. However, for a target of general convex shape the interval $[0,2\pi)$ for the phase $\xi_1$ can be covered several times depending also on the profile of the incident beam. If different profiles of the beam, with certain phase or velocity distributions are used, it is interesting to investigate the imprints that time-dependence leaves on the cross sections. The understanding gained by the present study opens up the perspective to conduct the inverse scattering problem for scattering off time-dependent targets. As shown this can be achieved by manipulation of the incident beam. With the appropriate tuning it is possible to extract information concerning the frequency, the amplitude and the axis of oscillation as well as the size of the scatterer. It is also interesting to study if the oscillation law, assumed to be harmonic in all the systems considered here, leaves its imprints in the peaks of the differential cross sections. Finally, it is of relevance to extend the present study to different setups supporting also the presence of UPOs in the scattering dynamics.

\section*{ACKNOWLEDGMENT} The authors thank Dr. P. K. Papachristou for helpful
discussions. L.B. acknowledges financial support from the projects IN-111607 (DGAPA-UNAM) and 79988 (CONACyT). P.S. acknowledges financial support by the Deutsche Forschungsgeeinschaft under the contract Schm 885/13.

\end{document}